\documentclass[twocolumn, aps, prd,showpacs, superscriptaddress]{revtex4-2}
\usepackage{temp}

\begin{document}
\title{
Analytical study of birefringent cavities for axion-like dark matter search}
\author{Tadashi Kuramoto}
    \email{tkuramot@post.kek.jp}
    \affiliation{Research Institute for Interdisciplinary Science, Okayama University, Okayama 700-8530, Japan}
    \affiliation{Theory Center, Institute for Particles and Nuclear Studies (IPNS), High Energy Accelerator Research Organization (KEK), Tsukuba, Ibaraki 305-0801, Japan}
    \affiliation{Graduate University for Advanced Studies (SOKENDAI), Hayama, Miura District, Kanagawa 240-0193, Japan}
\author{Yasutaka Imai}
    \affiliation{Research Institute for Interdisciplinary Science, Okayama University, Okayama 700-8530, Japan}
\author{Takahiko Masuda}
    \email{masuda@okayama-u.ac.jp}
    \affiliation{Research Institute for Interdisciplinary Science, Okayama University, Okayama 700-8530, Japan}
\author{Yutaka Shikano}
    \email{yshikano@cs.tsukuba.ac.jp}
    \affiliation{Institute of Systems and Information Engineering, University of Tsukuba, Tsukuba, Ibaraki 305-8573, Japan}
    \affiliation{Center for Artificial Intelligence Research, University of Tsukuba, Tsukuba, Ibaraki 305-8577, Japan}
    \affiliation{Institute for Quantum Studies, Chapman University, Orange, CA 92866, USA}
\author{Sayuri Takatori}
    \affiliation{Research Institute for Interdisciplinary Science, Okayama University, Okayama 700-8530, Japan}
\author{Satoshi Uetake}
    \email{uetake@okayama-u.ac.jp}
    \affiliation{Research Institute for Interdisciplinary Science, Okayama University, Okayama 700-8530, Japan}

\begin{abstract}

Light polarization plays a crucial role in optical-cavity experiments; however, mirror birefringence presents a significant challenge that must be addressed carefully. In this study, a rigorous, nonperturbative framework is developed to quantify birefringence effects by incorporating variations in reflectance and polarization misalignment. We analyze the impact of this framework on the sensitivity of axion-like particle (ALP) dark-matter searches. The results show that both birefringence and misalignment contribute to sensitivity degradation in the low-mass regime; however, the adverse effects of misalignment can be mitigated by selecting a postselection angle greater than the misalignment angle. Furthermore, birefringence produces an additional resonance peak in the high-mass region, which remains largely unaffected by misalignment and postselection variations. This rigorous framework underscores the importance of considering birefringence in high-precision optical-cavity experiments for ALP detection.
\end{abstract}
\maketitle

\section{Introduction}

Optical cavities are widely used in various fields, ranging from fundamental science to precision metrology \cite{YE20031}. They are commonly employed to generate coherent light and detect interference. 
Although lasers are indispensable tools for spectroscopy, the incorporation of optical cavities can significantly improve spectroscopy sensitivity and resolution, as demonstrated by techniques such as cavity-enhanced spectroscopy \cite{CESS,Wang2023}.
One of the most advanced applications of optical cavities is gravitational-wave detection, which uses large Fabry--P\'erot--Michelson interferometers \cite{Accadia_2012,Aasi2015,Akutsu2021}.
Recently, optical cavities have attracted attention in the search for unknown elementary particles that may be dark matter \cite{Derocco2018,Obata2018,Liu2019,Nagano2019,Nagano2021}. These are called axion-like particles (ALPs), which are ultralight pseudoscalar particles. If ALPs interact weakly with linearly polarized electromagnetic fields, they are expected to induce a small rotation in polarization. Because this effect is extremely small, optical cavities are used to amplify it. Several groups have reported experimental results in this context, including DANCE \cite{Oshima2023}, LIDA \cite{Heinze2024}, and ADBC \cite{Pandey2024}.

The high-finesse optical cavities used in such precise polarization measurements require careful consideration of the birefringence of optical components such as mirrors. Birefringence within the cavity can cause polarization degradation, which becomes non-negligible in high-finesse systems. Because ALP searches rely on precise polarization measurements, birefringence may be a critical limiting factor.

Similar experiments focusing on polarization measurements with optical cavities have been conducted to investigate vacuum magnetic birefringence \cite{Cadene2014,Fan2018,Ejlli2020}. A critical issue in these measurements is the imperfection of the mirrors comprising the optical cavity, particularly mirror birefringence and its fluctuations. Therefore, these effects have been extensively studied \cite{Hartman2017,Ejlli2020,Agil2022}.
The optical cavities used in vacuum magnetic-birefringence measurements are Fabry--P\'erot cavities, in which a pair of mirrors is placed parallel to the optical path and the light-reflection angle is normal to the mirror surface. 
Mirror-birefringence effects in Fabry--P\'erot cavities have also been investigated in gravitational-wave detectors. Gravitational-wave experiments require extremely high sensitivity; thus, even the birefringence inhomogeneity on a mirror surface may affect the experiment sensitivity \cite{Michimura2024,Tanioka2023,Wang2024}.

In contrast, optical cavities used for ALP dark-matter searches typically require more complex configurations than Fabry--P\'erot cavities. The polarization rotation induced by the ALP field is effectively canceled at each reflection in a cavity with equal path lengths between adjacent mirrors. Therefore, ring cavities with asymmetric optical paths are typically employed. In such ring cavities, the angle of incidence at the mirrors is non-normal, which introduces a significant birefringence effect because of the phase difference between the s- and p-polarized components. Some experimental setups exploit this phase difference to enhance sensitivity to ALPs in specific mass ranges \cite{Heinze2024,Pandey2024}.

Despite the higher sensitivity to birefringence in ring-cavity configurations, the effect of birefringent mirrors on ALP searches has not been thoroughly investigated. In this study, we investigate the influence of mirror birefringence on polarization-based ALP dark-matter searches using rigorous analysis. Section~\ref{sec:cavity} discusses the detailed dynamics of the electromagnetic field in a cavity using birefringent mirrors.
An analytical solution for the polarization evolution in the presence of an ALP field is derived as described in Section~\ref{sec:cavity_in_the_ALP_field}.

In Section~\ref{sec:discussion}, we discuss the implications of the birefringence effect on the sensitivity of ALP searches, and we present our conclusions in Section~\ref{sec:conclusion}. Detailed algebraic derivations to improve readability are provided in the Appendix. In Appendix A, we derive the time-evolution operator for polarized light in the ALP field. Appendix B provides the complete expressions of (\ref{eq: bmC^pm_f}). In Appendix C, we discuss the case in which birefringence does not change the linear polarization. In Appendix D, we provide an approximate resonance condition when birefringence does not split, but slightly modifies the spectrum. Appendix E provides the relationship between our parameters and measurable quantities. In Appendix F, we propose a cavity that suppresses birefringence.

\section{Cavity with Birefringence\label{sec:cavity}}

\begin{figure}[H]
    \begin{tabular}{cc}
        \multicolumn{2}{c}{
        \begin{minipage}[b]{1.0\linewidth}
            \centering
            \includegraphics[width=1.0\linewidth, bb = 130 100 1720 1080]{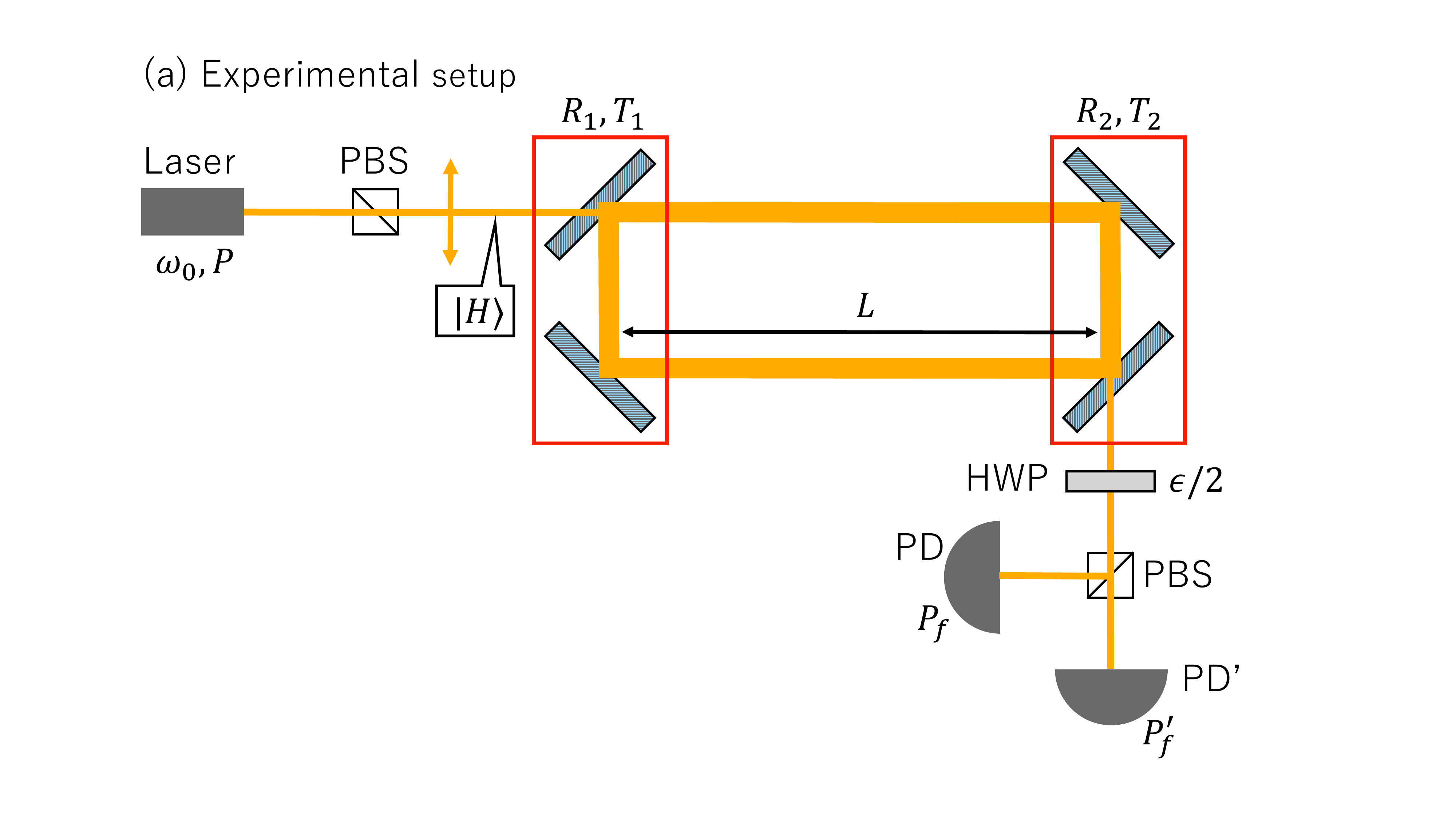}
        \end{minipage}} \\
        \multicolumn{2}{c}{
        \begin{minipage}{1.0\linewidth}
            \centering
            \includegraphics[width=1.0\linewidth, bb = 0 270 960 540]{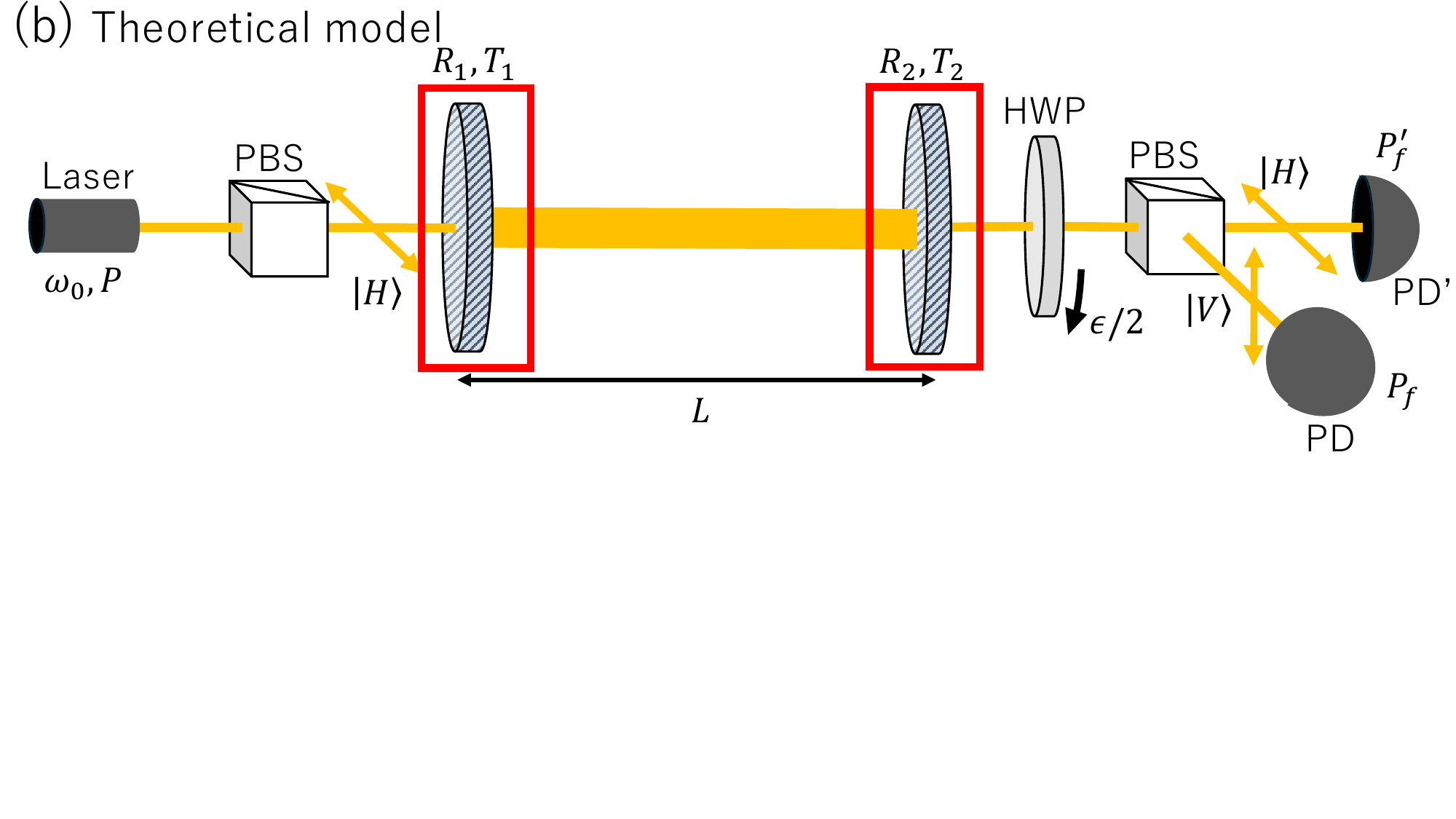}
        \end{minipage}} \\
        \begin{minipage}{0.5\linewidth}
            \centering
            \includegraphics[width=1.05\linewidth, bb = 0 0 842 595]{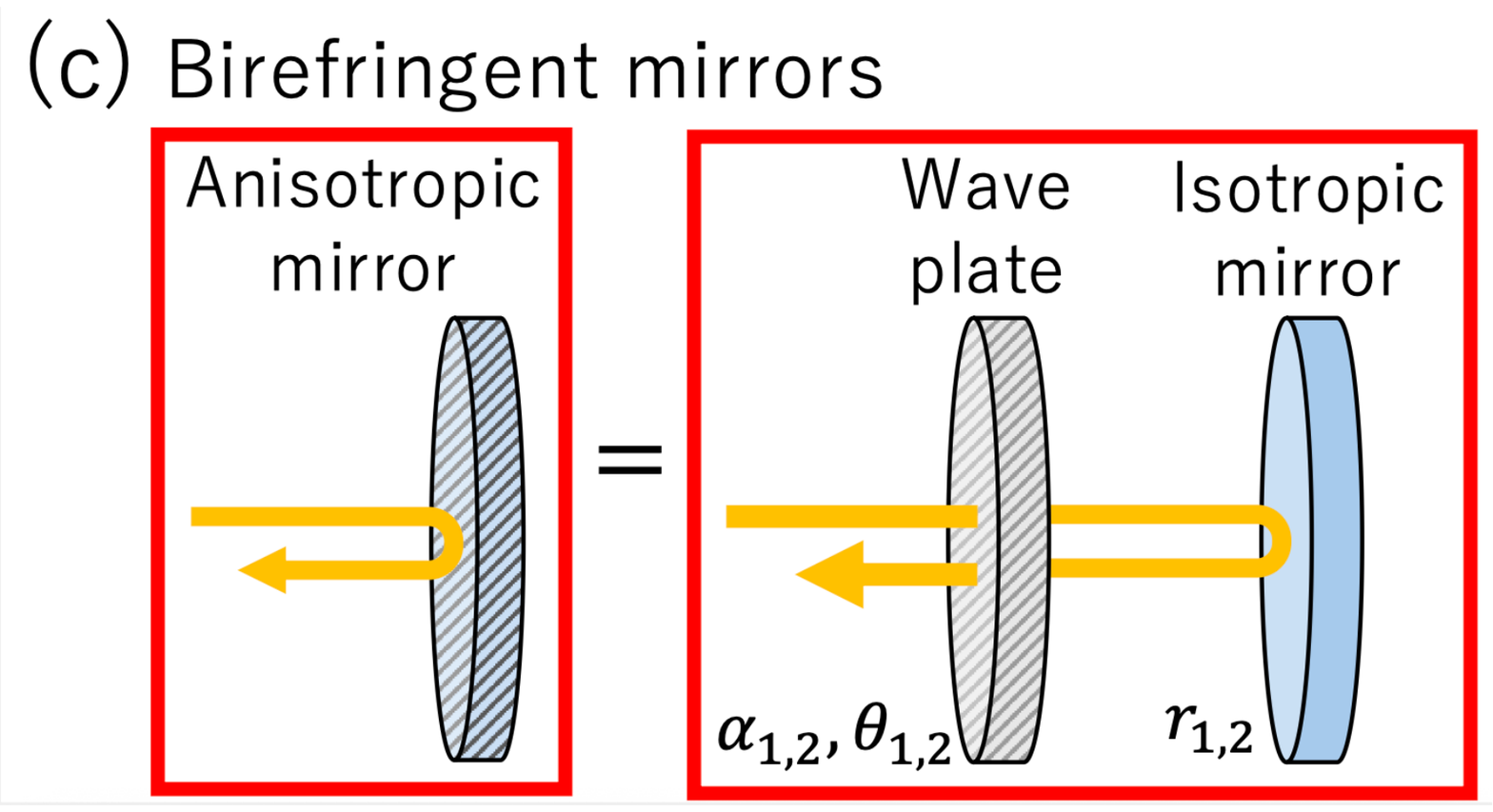}
        \end{minipage} &
        \begin{minipage}{0.5\linewidth}
            \centering
            \includegraphics[width=0.9\linewidth, bb = 0 0 842 595]{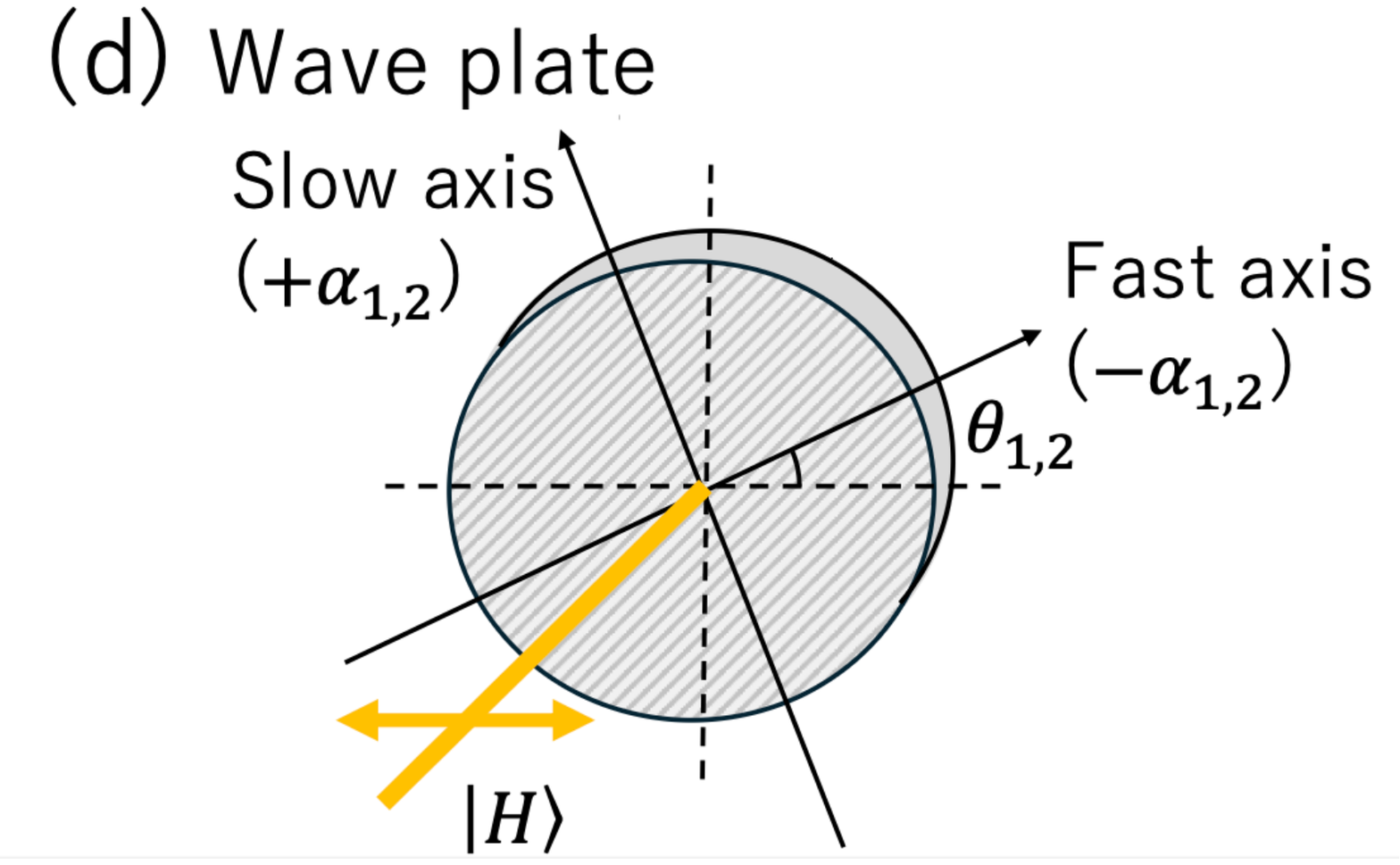}
        \end{minipage}
    \end{tabular}
    \caption{(a) Schematic of the experimental setup considered in this work. $\omega_0$, angular frequency of laser light; $P$, laser power; $|H\rangle$, horizontal linear polarization; $R_{1,2}~(T_{1,2})$, Jones matrices of reflection (transmission) representing the two successive mirrors; $L$, cavity length; $\epsilon$, rotation angle of linear polarization; PBS, polarizing beam splitter; HWP, half-wave plate; PD, photodetector. (b) Schematic of the theoretical setup considered in this work. The two successive mirrors at the shorter sides of the ring cavity are regarded as a single birefringent mirror. (c) Model of a birefringent mirror. $r_{1,2}$, (geometric) average of reflectances along the fast and slow axes of the waveplate. (d) Waveplates characterized by retardation $\alpha_{1,2}$, and angle between the $|H\rangle$ polarization and its fast axes $\theta_{1,2}$.}
    \label{fig: Fabry-Perot}
\end{figure}

A ring cavity is typically employed in ALP search using an optical cavity. In a simple Fabry--P\'erot cavity, the effect of the ALP field is canceled after the light makes a round trip because the direction of the polarization change caused by the ALP field is reversed upon a single reflection. In this study, we assume a ring cavity with either a rectangular or bowtie shape, where the longer sides are significantly longer than the shorter ones (Fig.~\ref{fig: Fabry-Perot}(a)). 

In this section, we describe development of an analytical model for the propagation of an electromagnetic field in a ring cavity composed of birefringent mirrors in the absence of an ALP field. A schematic of the optical setup considered in this section is shown in Fig.~\ref{fig: Fabry-Perot}(a). The dominant component of the birefringence originates from the oblique incidence of light in the mirrors in this configuration. 
The laser emits light with angular frequency $\omega_0$ and power $P$. The light is cleaned by the first polarizing beam splitter (PBS) to be linearly polarized $|H\rangle$ and then enters a rectangular cavity of length $L$. After multiple reflections inside the cavity, the transmitted light exits and passes through a half-wave plate (HWP), which rotates the polarization by an angle $\epsilon$. The second PBS splits the light into orthogonal polarization components. Finally, the two photodetectors (PDs) measure the power of each polarization component $P_f$ and $P'_f$. $P_f (P'_f)$ refers to the vertically (horizontally) polarized light.

 We consider a ring cavity with a high aspect ratio in which the two reflections at the short sides can be treated as a single effective reflection and transmission (Fig.~\ref{fig: Fabry-Perot}(b)). The Jones matrices for the reflection and transmission of the input (output) mirror are $R_1(R_2)$ and $T_1(T_2)$, respectively. A model of a birefringent mirror is shown in Fig.~\ref{fig: Fabry-Perot}(c). A mirror with birefringence is represented by an isotropic mirror without birefringence that is covered with a waveplate. A model of the waveplate is shown in Fig.~\ref{fig: Fabry-Perot}(d). To describe the effect of mirror birefringence, we assume that the waveplate has a complex retardation $\alpha_{1,2}$ for each axis and is rotated along the polarization plane at an angle of $\theta_{1,2}$. The Jones matrix $R_j~(j=1,2)$ is written as
\begin{align}
    \label{eq: reflection matrix}
    R_j&=r_je^{-i\theta_j\sigma_y}e^{-i\alpha_j\sigma_z}e^{i\theta_j\sigma_y} \\
 \begin{split}
    \label{eq: reflection matrix linear to Pauli}
    &=r_j\Bigl\{\cos{\alpha_j} \\
    &\hspace{30pt}-i\sin{\alpha_j}\left[\sigma_z\cos{(2\theta_j)}+\sigma_x\sin{(2\theta_j)} \right]\Bigr\},
 \end{split}
\end{align}
where $r_j\in\mathbb{R}$ is the isotropic component of the amplitude reflectance, $\alpha_j\in\mathbb{C}$ is the complex retardation, $\theta_j\in\mathbb{R}$ is the angle between the horizontal polarization of light and fast axis of the waveplate, and $\sigma_{x,y,z}$ are the Pauli matrices. The real part of $\alpha_j$ is the phase retardation, and the imaginary part of $\alpha_j$ represents the reflectance anisotropy. The actual reflectances for linearly polarized light along the fast and slow axes of the waveplate are $r_je^{\pm\mathrm{Im}[\alpha_j]}$. Similarly, the Jones matrices for transmission $T_j~(j=1,2)$ are expressed as
\begin{align}
    \label{eq: transmission matrix}
    T_j&=t_je^{-i\theta_j\sigma_y}e^{\beta_j\sigma_z}e^{i\theta_j\sigma_y} \\
 \begin{split}
    &=t_j\Bigl\{\cosh{\beta_j} \\
    &\hspace{30pt}+\sinh{\beta_j}\left[\sigma_z\cos{(2\theta_j)}+\sigma_x\sin{(2\theta_j)} \right]\Bigr\},
 \end{split}
\end{align}
where $t_j\in\mathbb{R}$ is the average transmittance and $\beta_j\in\mathbb{R}$ is the imaginary part of the complex retardation. We assume that the real part of the complex retardation can be ignored because the light does not transmit through the mirror many times. Moreover we assume that the angle of the waveplate of transmission is the same as that of the waveplate of reflection. Owing to energy conservation, the parameters $t_j$ and $\beta_j$ are written using $r_j$ and $\alpha_j$ as
\begin{align}
    t_j&=\left[(1-r_j^2)^2-4r_j^2\sinh^2(\mathrm{Im}[\alpha_j])\right]^{1/4},
\end{align}
and
\begin{align}
    \label{eq: beta_j}
    \beta_j&=\frac{1}{4}\log{\left(\frac{1-r_j^2e^{2\mathrm{Im}[\alpha_j]}}{1-r_j^2e^{-2\mathrm{Im}[\alpha_j]}}\right)}.
\end{align}

The calculations presented in this section follow a method similar to that of \cite{Brandi1997}. In this previous study, the retardations $\alpha_{1,2}$ were assumed to be real and small; that is $\alpha_{1,2}\in \mathbb{R},~\alpha_{1,2}\ll 1$. Our analysis is fully nonperturbative and valid for complex values of $\alpha_{1,2}$. Notably, the values of $\alpha_{1,2}$ are bounded; its real part is compact owing to the periodicity of the waveplate:
\begin{align}
    -\pi\leq\mathrm{Re}\left[\alpha_j\right]\leq\pi~~~(j=1,2).
\end{align}
The imaginary part is constrained by the physical requirement that the reflectance does not exceed unity:
\begin{align}
    r_je^{\left|\mathrm{Im}\left[\alpha_j\right]\right|}\leq1~~~(j=1,2),
\end{align}
or
\begin{align}
    \label{eq: Im alpha limitation}
    -\ln{\left(\frac{1}{r_j}\right)}\leq\mathrm{Im}\left[\alpha_j\right]\leq\ln{\left(\frac{1}{r_j}\right)}~~~(j=1,2).
\end{align}
 If $r_1=r_2$ and the averaged finesse $\mathcal{F}=\pi\sqrt{r_1r_2}/(1-r_1r_2)=\pi r_1/(1-r_1^2)$ is high, the condition for $\mathrm{Im}\left[\alpha_{1,2}\right]$ can be approximated as
 \begin{align}
     \label{eq: Im alpha limitation approx}
     -\frac{\pi}{2\mathcal{F}}\lesssim\mathrm{Im}\left[\alpha_{1,2}\right]\lesssim\frac{\pi}{2\mathcal{F}},
 \end{align}
which implies that the values of $\mathrm{Im}\left[\alpha_{1,2}\right]$ must be small for a high-finesse cavity.

When the initial state of light is horizontally polarized with angular frequency $\omega_0$, the electromagnetic field throughout the output mirror is
\begin{align}
    \bm{A}_f(t;\omega_0)&=A_0T_2\bm{C}^{(0)}_fe^{-i\omega_0t}, \\
    \bm{C}^{(0)}_f&:=\sum_{n=0}^{\infty}\bm{C}^{(0)}_n, \\
    \bm{C}^{(0)}_n&=e^{2i\omega_0L}R_1R_2\bm{C}^{(0)}_{n-1}, \\
    \bm{C}^{(0)}_0&=e^{i\omega_0L}T_1\begin{pmatrix} 1 \\ 0 \end{pmatrix}.
\end{align}
The $n$th reflected component $\bm{C}^{(0)}_n~(n\geq0)$ is given by
\begin{align} \bm{C}^{(0)}_n=e^{i\omega_0L}\left(r_1r_2e^{2i\omega_0L}\right)^n\Gamma^nT_1\begin{pmatrix} 1 \\ 0 \end{pmatrix},
\end{align}
using the Jones matrix of the combined waveplate:
\begin{align}
    \Gamma&:=\frac{1}{r_1r_2}R_1R_2.
\end{align}
The matrix $\Gamma$ can be diagonalized as
\begin{align}
    \label{eq:Gamma diagonalized}
    \Gamma=e^{-i\theta\bm{n}\cdot\bm{\sigma}}e^{-i\alpha\sigma_z}e^{i\theta\bm{n}\cdot\bm{\sigma}},
\end{align}
where $\bm{n}$ is a vector parallel to the rotation axis and is expressed as
\begin{align}
    \bm{n}:=\begin{pmatrix} -\sin{\phi} \\ \cos{\phi} \\ 0 \end{pmatrix},
\end{align}
and $\alpha,\theta,\phi$ are functions of $\alpha_{1,2}$ and $\theta_{1,2}$:
\begin{align}
    \label{eq: alpha}
    \cos{\alpha}&:=\cos{\alpha_1}\cos{\alpha_2}-\sin{\alpha_1}\sin{\alpha_2}\cos{\left[2\left(\theta_1-\theta_2\right)\right]}, \\
    \label{eq: theta}
    \cos{\left(2\theta\right)}&:=\frac{\sin{\alpha_1}\cos{\alpha_2}\cos{\left(2\theta_1\right)}+\cos{\alpha_1}\sin{\alpha_2}\cos{\left(2\theta_2\right)}}{\sin{\alpha}},
\end{align}
and
\begin{align}
    \label{eq: phi}
    \tan{\phi}&:=\frac{-\sin{\alpha_1}\sin{\alpha_2}\sin{\left[2\left(\theta_1-\theta_2\right)\right]}}{\sin{\alpha_1}\cos{\alpha_2}\sin{\left(2\theta_1\right)}+\cos{\alpha_1}\sin{\alpha_2}\sin{\left(2\theta_2\right)}}.
\end{align}
Matrix $e^{-i\alpha\sigma_z}$ in (\ref{eq:Gamma diagonalized}) is diagonal with eigenvalues $e^{\mp i\alpha}$. This diagonalization defines a combined waveplate with an effective retardation $\alpha$ and fast-axis angle $\theta$. $\Gamma^n$ is expressed by the eigenvalues to the power of $n$, and $\bm{C}^{(0)}_f$ is reduced using $\alpha$, $\theta$, and $\bm{n}$:
\begin{align}
 \label{eq: C^0_f}
 \begin{split}
    \bm{C}^{(0)}_f&=e^{i\omega_0L}e^{-i\theta\bm{n}\cdot\bm{\sigma}} \\
    &\hspace{10pt}\times\begin{pmatrix} \frac{1}{1-r_1r_2e^{2i\omega_0L}e^{-i\alpha}} & 0 \\ 0 & \frac{1}{1-r_1r_2e^{2i\omega_0L}e^{i\alpha}} \end{pmatrix} \\
    &\hspace{10pt}\times e^{i\theta\bm{n}\cdot\bm{\sigma}}T_1\begin{pmatrix} 1 \\ 0 \end{pmatrix}.
 \end{split}
\end{align}
This expression indicates that if the input polarization is aligned with one of the fast or slow axes of the combined waveplate in the cavity, then the resulting output intensity is proportional to
\begin{align}
    \left|\frac{1}{1-r_1r_2e^{2i\omega_0L}e^{-i\alpha}}\right|^2
\end{align}
for the fast axis or
\begin{align}
    \left|\frac{1}{1-r_1r_2e^{2i\omega_0L}e^{i\alpha}}\right|^2
\end{align}
for the slow axis, which are the two Lorentzians with different peak positions and widths. For a general polarization, we observe their superpositions. The resonance conditions determining the positions of the two peaks are
\begin{align}
 \label{eq: resonance condition}
    2\omega_0L\pm\mathrm{Re}[\alpha]=2\pi l~~~(l\in\mathbb{Z}).
\end{align}
The two finesses that determine the widths of the two peaks are
\begin{align}
    \label{eq: actual finesses}
    \mathcal{F}_{F/S}:=\frac{\pi\sqrt{r_1r_2e^{\pm\mathrm{Im}\left[\alpha\right]}}}{1-r_1r_2e^{\pm\mathrm{Im}\left[\alpha\right]}}.
\end{align}
The real part of $\alpha$ shifts and splits the peaks of the resonance curve, whereas the imaginary part of $\alpha$ makes one of the peaks sharper and the other broader. If the angle between the waveplates at the mirrors is zero, that is, $\theta_1=\theta_2=0$, the angle of the combined waveplate is the same as that of the mirrors, $\theta=\theta_1=\theta_2$, and the retardation is simply the sum of the retardations of each waveplate, $\alpha=\alpha_1+\alpha_2$.

The power of the light is observed during experiments. Generally, we can choose a linear polarization with a particular postselection angle $\epsilon$:
\begin{align}
    \bm{n}'_f:=\begin{pmatrix} \cos{\epsilon} \\ -\sin{\epsilon} \end{pmatrix}.
\end{align}
The postselection angle $\epsilon$ is a controllable parameter and we assume that $\epsilon$ is small. The observed power is
\begin{align}
 \label{eq: P'_f}
 \begin{split}
    P'_f(t;\omega_0)&=P\Bigl|\frac{a'_F}{1-r_1r_2e^{2i\omega_0L}e^{-i\alpha}} \\
    &~~~~~~~~~~~~~~~~+\frac{a'_S}{1-r_1r_2e^{2i\omega_0L}e^{i\alpha}}\Bigr|^2,
 \end{split}
\end{align}
where $P$ is the laser power before entering the cavity, and
\begin{align}
    a'_{F/S}:=\bm{n}'_f\cdot T_2e^{-i\theta\bm{n}\cdot\bm{\sigma}}\frac{1\pm\sigma_z}{2}e^{i\theta\bm{n}\cdot\bm{\sigma}}T_1\begin{pmatrix} 1 \\ 0 \end{pmatrix}
\end{align}
indicates how much of the polarizations along the fast and slow axes of the combined waveplate represented by $\Gamma$ in (\ref{eq:Gamma diagonalized}) exist within the cavity.

Similar to $P'_f$, the power in the other port is
\begin{align}
 \label{eq: P'_f}
 \begin{split}
    P_f(t;\omega_0)&=P\Bigl|\frac{a_F}{1-r_1r_2e^{2i\omega_0L}e^{-i\alpha}} \\
    &~~~~~~~~~~~~~~~~+\frac{a_S}{1-r_1r_2e^{2i\omega_0L}e^{i\alpha}}\Bigr|^2,
 \end{split}
\end{align}
where
\begin{align}
    a_{F/S}:=\bm{n}_f\cdot T_2e^{-i\theta\bm{n}\cdot\bm{\sigma}}\frac{1\pm\sigma_z}{2}e^{i\theta\bm{n}\cdot\bm{\sigma}}T_1\begin{pmatrix} 1 \\ 0 \end{pmatrix},
\end{align}
and
\begin{align}
    \label{eq: n_f}
    \bm{n}_f:=\begin{pmatrix} \sin{\epsilon} \\ \cos{\epsilon} \end{pmatrix}
\end{align}
is the other postselected polarization vector, which is orthogonal to $\bm{n}'_f$. When $\alpha_{1,2}$ and $\theta_{1,2}$ are small, the power $P_f$ becomes small relative to $P'_f$ because the polarization $\bm{n}_f$ with a small $\epsilon$ is almost orthogonal to the initial polarization.

$P_f/(P_f+P'_f)$ is used to evaluate the population of light that becomes vertically polarized throughout the process. No population of light is transferred when the retardation $\alpha_1$ or the angle of the birefringence axis $\theta_1$ is zero, except for a slight rotation of the polarization at the HWP, as shown in Fig.~\ref{fig: A vs alpha,theta}. As shown in Fig.~\ref{fig: A vs alpha,theta}(a), the population transfer starts to increase when $\alpha_1\sim1/\mathcal{F}$ and saturates when $\alpha_1\gg1/\mathcal{F}$ with values of almost $\sin^2{\theta_1}$, as also shown in Fig.~\ref{fig: A vs alpha,theta}(b). When $\alpha_1\gg1/\mathcal{F}$ and $\theta_1\neq0$, the resonance curve exhibits two divided peaks corresponding to the two resonance conditions in (\ref{eq: resonance condition}).

\begin{figure}[H]
 \begin{tabular}{cc}
    (a) &  \\
     & \begin{minipage}[b]{0.9\linewidth}
         \centering
    \includegraphics[width=1.0\linewidth, bb = 0 0 360 233]{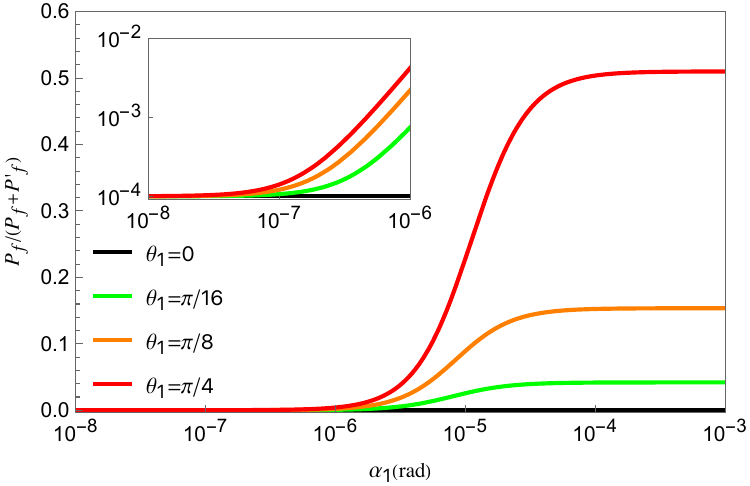}
     \end{minipage} \\
     (b) &  \\
      & \begin{minipage}[b]{0.9\linewidth}
          \centering
    \includegraphics[width=1.0\linewidth, bb = 0 0 360 238]{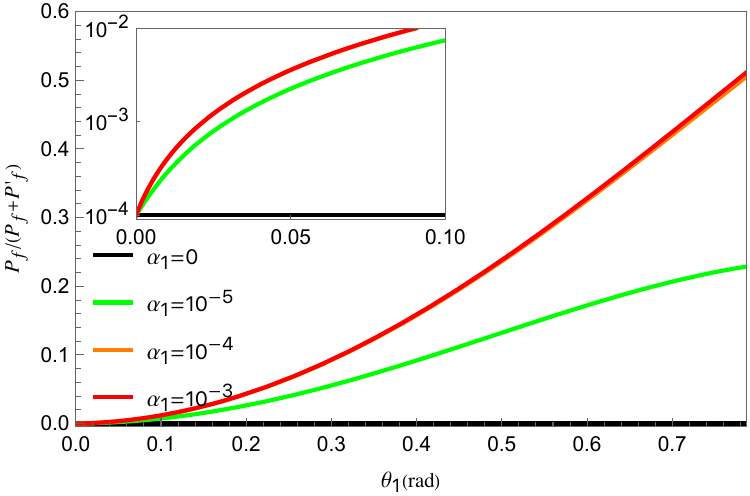}
      \end{minipage} 
 \end{tabular}
   \caption{Population transfer with the real retardation $\alpha_1$ and angle of birefringence axis $\theta_1$. $P_f~(P'_f)$ is the laser power at the detection (reference) port, in which the polarization is orthogonal (parallel) to the polarization of light just before entering the cavity. The minimum value $\sin^2{\epsilon}\simeq\epsilon^2$ is due to the postselection. Common parameters are $\mathcal{F}=10^5~(r_1=r_2)$, $L=10.64~\mathrm{m}$, $\lambda=1064~\mathrm{nm}$, $\alpha_2=\alpha_1$, $\theta_2=\theta_1$, and $\epsilon=0.01$. The resonance condition in (\ref{eq: resonance condition theta_1,2=0}) holds.}
    \label{fig: A vs alpha,theta}
\end{figure}

The overall reflectance of the cavity $[1-(P_f+P'_f)/P]$ also depends on $\alpha_1$ and $\theta_1$. The sum of the output powers $(P_f+P'_f)$ decreases as the retardation $\alpha_1$ and/or the angle of the birefringence axis $\theta_1$ increases, because in either situation, more light flows into the off-resonant polarization mode within the cavity. As shown in Fig.~\ref{fig: B vs alpha,theta}(a), the dependence of the reflectance on $\alpha_1$ is similar to that of the polarization transfer shown in Fig.~\ref{fig: A vs alpha,theta}(a); however, the slope is slightly steeper. Fig.~\ref{fig: B vs alpha,theta}(b) shows that the reflectance is always proportional to $\sin^2{\theta_1}$, in contrast to the more nonlinear behavior shown in Fig.~\ref{fig: A vs alpha,theta}(b). These behaviors will be modified when the ALP effect is introduced.

\begin{figure}[H]
 \begin{tabular}{cc}
    (a) &  \\
     & \begin{minipage}[b]{0.9\linewidth}
         \centering
    \includegraphics[width=1.0\linewidth, bb = 0 0 360 233]{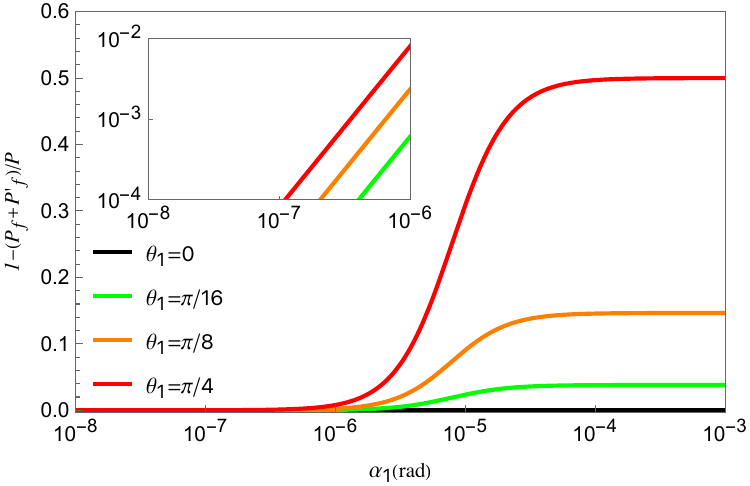}
     \end{minipage} \\
     (b) &  \\
      & \begin{minipage}[b]{0.9\linewidth}
          \centering
    \includegraphics[width=1.0\linewidth, bb = 0 0 360 238]{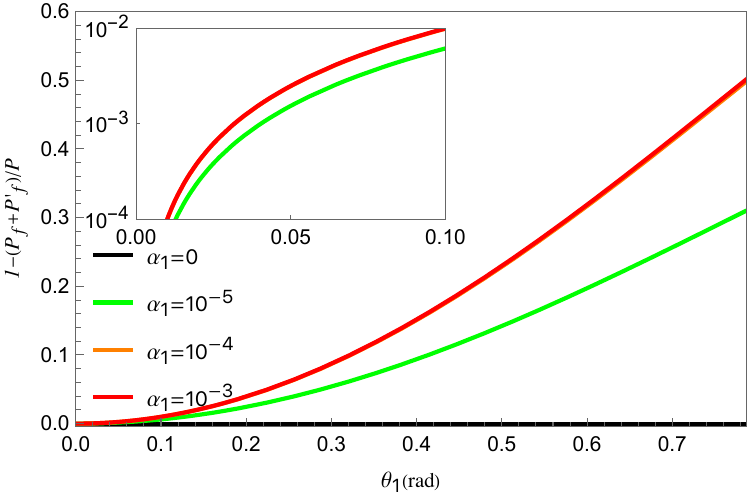}
      \end{minipage}
 \end{tabular}
    \caption{Reflectance of the cavity with the real retardation $\alpha_1$ and angle of birefringence axis $\theta_1$. $P$ is the laser power just before entering the cavity. $P_f~(P'_f)$ is the laser power at the detection (reference) port, in which the polarization is orthogonal (parallel) to the polarization of light before entering the cavity. Common parameters are $\mathcal{F}=10^5~(r_1=r_2)$, $L=10.64~\mathrm{m}$, $\lambda=1064~\mathrm{nm}$, $\alpha_2=\alpha_1$, $\theta_2=\theta_1$, and $\epsilon=0.01$. The resonance condition in (\ref{eq: resonance condition theta_1,2=0}) holds.}
    \label{fig: B vs alpha,theta}
\end{figure}

\section{Cavity in the ALP field\label{sec:cavity_in_the_ALP_field}}

We introduce the ALP effect into the developed model presented in the previous section. The elementary process of the interaction between the ALP and light fields has been studied \cite{Derocco2018,Obata2018,Liu2019,Martynov2020}, and the derivation is summarized in Appendix A.

The effect of the ALP field on the shorter sides is ignored, and the cavity is treated in the same manner as in Section~\ref{sec:cavity}, except for the presence of the ALP field in the longer sides of the ring cavity.

In the presence of an ALP field, two sidebands with a frequency $\omega_0\pm m_a$ emerge in the electromagnetic field through the opposite mirror, where $m_a$ is the mass of the ALP. The modification in the calculation involves operating the time-evolution operator $U$ in (\ref{eq: ALP operator}) during light propagation within the cavity. The recurrence relation of the light in the cavity is
\begin{align}
 \label{eq: recurrence light}
 \begin{split}
    \bm{A}_n(t;\omega_0)&=U(t,t-L)R_1U(t-L,t-2L) \\
    &~~~~~~\times R_2\bm{A}_{n-1}(t-2L;\omega_0),
 \end{split}
\end{align}
where $L$ is the cavity length in the long side (Fig.~\ref{fig: Fabry-Perot}(a)) and
\begin{align}
 \begin{split}
    \bm{A}_n(t;\omega_0)&=A_0T_2\Bigl[ \bm{C}^{(0)}_n+\bm{C}^{(+)}_ne^{-i(m_at+\delta_a)} \\
    &~~~~~~~~~~~~~~~~~~+\bm{C}^{(-)}_ne^{i(m_at+\delta_a)} \Bigr]e^{-i\omega_0t}.
 \end{split}
\end{align}
Equation (\ref{eq: recurrence light}) results in the following recurrence relations for each coefficient:
\begin{align}
    \bm{C}^{(0)}_n&=r_1r_2e^{i\omega_0L}\Gamma\bm{C}^{(0)}_{n-1}, \\
 \begin{split}
    \bm{C}^{(\pm)}_n&=r_1r_2e^{i(\omega_0\pm m_a)L}\Gamma\bm{C}^{(\pm)}_{n-1} \\
    &~~~-\frac{i}{2}g\left(1-e^{\pm im_aL}\right)e^{\mp i\delta_a}\Xi_\pm\bm{C}^{(0)}_n,
 \end{split}
\end{align}
where $g$ is given by (\ref{eq: perturbation parameter}), $\delta_a$ is the constant phase of the ALP field, and the matrix $\Xi_\pm$ is defined as
\begin{align}
    \label{eq: Xi_pm}
    \Xi_\pm&:=\sigma_y+e^{\pm im_aL}R_1\sigma_yR^{-1}_1.
\end{align}
The initial conditions are
\begin{align}
    \bm{C}^{(0)}_0&=e^{i\omega_0L}T_1\begin{pmatrix} 1 \\ 0 \end{pmatrix},
\end{align}
and
\begin{align}
    \bm{C}^{(\pm)}_0&=e^{i\omega_0L}\frac{1}{2}g\left(1-e^{\pm im_aL}\right)e^{\mp i\delta_a}\left(-i\sigma_y\right)T_1\begin{pmatrix} 1 \\ 0 \end{pmatrix}.
\end{align}
The final state of the light is
\begin{align}
    \bm{A}_f(t;\omega_0)&=\sum_{n=0}^\infty\bm{A}_n(t;\omega_0) \\
 \label{eq: A_f=C_f^(0,+,-)}
 \begin{split}
    &=A_0T_2\Bigl[ \bm{C}^{(0)}_f+\bm{C}^{(+)}_fe^{-i(m_at+\delta_a)} \\
    &~~~~~~~~~~~~~~~~~~+\bm{C}^{(-)}_fe^{i(m_at+\delta_a)} \Bigr]e^{-i\omega_0t},
 \end{split}\\
 \bm{C}^{(0,\pm)}_f&:=\sum_{n=0}^{\infty}\bm{C}^{(0,\pm)}_n. 
\end{align}
In the linear approximation with respect to $g$, $\bm{C}^{(0)}_f$ is the same as (\ref{eq: C^0_f}), and $\bm{C}^{(\pm)}_f$ are
\begin{align}
 \label{eq: bmC^pm_f}
 \begin{split}
    \bm{C}^{(\pm)}_f&=\frac{1}{2}ge^{i\omega_0L}\left(1-e^{\pm im_aL}\right) \\
    &\times \Bigl\{ \frac{\bm{u}^{(\pm)}_F}{1-r_1r_2e^{2i(\omega_0\pm m_a)L}e^{-i\alpha}} \\
    &~~~~+\frac{\bm{u}^{(\pm)}_S}{1-r_1r_2e^{2i(\omega_0\pm m_a)L}e^{i\alpha}} \\
    &~~~~+\frac{\bm{v}^{(\pm)}_F}{1-r_1r_2e^{2i\omega_0L}e^{-i\alpha}}+\frac{\bm{v}^{(\pm)}_S}{1-r_1r_2e^{2i\omega_0L}e^{i\alpha}} \Bigr\},
 \end{split}
\end{align}
where $\bm{u}^{(\pm)}_{F/S}$ and $\bm{v}^{(\pm)}_{F/S}$ are expressed by (\ref{eq:  u^pm_F}--\ref{eq: v^pm_S}). The sideband amplitude $\bm{C}_f^{(\pm)}$ in (\ref{eq: bmC^pm_f}) is the superposition of the four resonance factors of both the carrier light and signal light:
\begin{align}
    &\frac{1}{1-r_1r_2e^{2i(\omega_0\pm m_a)L}e^{-i\alpha}}, \\
    &\frac{1}{1-r_1r_2e^{2i(\omega_0\pm m_a)L}e^{i\alpha}}, \\
    &\frac{1}{1-r_1r_2e^{2i\omega_0L}e^{-i\alpha}},
\end{align}
and
\begin{align}
    &\frac{1}{1-r_1r_2e^{2i\omega_0L}e^{i\alpha}}.
\end{align}
The last two factors are present in (\ref{eq: C^0_f}) corresponding to the resonance of two carrier-light polarizations. The first two factors originate from the resonance of the signal light, which is modified by the birefringence of the cavity mirrors under the resonance condition:
\begin{align}
    \label{eq: sideband resonance}
    2\left(\omega_0\pm m_a\right)L\pm\mathrm{Re}\left[\alpha\right]=2\pi l~~~\left(l\in\mathbb{Z}\right).
\end{align}
The difference between the four modes can be interpreted in terms of the order of transitions between the carrier and signal modes, where the resonance factor of the carrier light arises when the field first resonates in the cavity and then transits to a sideband by ALP interaction, whereas the resonance factor of the signal corresponds to the reverse process.

As described in Section~\ref{sec:cavity}, we choose the output polarization $\bm{n}_f$ in (\ref{eq: n_f}) because the ALP field changes the polarization of light and the signal appears mainly in this port, whereas the other port is used to stabilize $\omega_0$ at the resonance frequency. The observed power $P_f$ is modified by the ALP field:
\begin{align}
 \begin{split}
    P_f(t;\omega_0)&=P\Bigl|\bm{n}_f\cdot T_2\bm{C}^{(0)}_f+\bm{n}_f\cdot T_2\bm{C}^{(+)}_fe^{-i(m_at+\delta_a)} \\
    &~~~~~~~~~~~~~~~~~~~~~~+\bm{n}_f\cdot T_2\bm{C}^{(-)}_fe^{i(m_at+\delta_a)}\Bigr|^2
 \end{split}\\
 \begin{split}
    &\simeq P_0+P_1\cos{\left(m_at+\delta_a-\chi \right)},
 \end{split}
\end{align}
where $P$ denotes the laser power before entering the cavity, and
\begin{align}
    P_0&:=P\left|C^{(0)}_f\right|^2, \\
    P_1&:=2P\left|C^{(0)*}_fC^{(+)}_f+C^{(0)}_fC^{(-)*}_f\right|, \\
    C^{(0,\pm)}_f&:=\bm{n}_f\cdot T_2\bm{C}^{(0,\pm)}_f,
\end{align}
where $P_0$ is the offset power. The root mean square of the second term, $P_1/\sqrt{2}$, is considered the signal power.

\section{Discussion\label{sec:discussion}}

The signal-to-noise ratio (SNR) for detecting the ALP-induced signal, limited by the shot noise of the offset power, is given by
\begin{align}
    \mathrm{SNR}&=\frac{P_1/\sqrt{2}}{\sqrt{S_{\mathrm{shot}}/\mathcal{T}_{\mathrm{obs}}}} \\
 \label{eq: SNR}
    &=\frac{\sqrt{2}}{\left|C^{(0)}_f\right|}\left|C^{(0)*}_fC^{(+)}_f+C^{(0)}_fC^{(-)*}_f\right|\sqrt{\frac{P\mathcal{T}_{\mathrm{obs}}}{\hbar\omega_0}},
\end{align}
where $S_{\mathrm{shot}}=\hbar\omega_0P_0$ denotes the power spectral density of the shot noise associated with the offset power $P_0$. The factor $\mathcal{T}_{\mathrm{obs}}$ represents the improvement in the $\mathrm{SNR}$ by the observation time, which depends on whether the observation time $T_{\mathrm{obs}}$ is larger than the coherent time of the ALP field $\tau$ \cite{Budker2014}:
\begin{align}
    \label{eq: time improvement factor}
    \mathcal{T}_\mathrm{obs}=\begin{cases} T_{\mathrm{obs}} & \left(T_\mathrm{obs}\lesssim\tau\right) \\ \left(T_\mathrm{obs}\tau\right)^{1/2} & \left(T_\mathrm{obs}\gtrsim\tau\right)  \end{cases},
\end{align}
For $T_{\mathrm{obs}}\lesssim\tau$, the SNR is proportional to the square root of $PT_{\mathrm{obs}}/\hbar\omega_0$.

$\mathrm{SNR}$ in (\ref{eq: SNR}) is proportional to the ALP-photon coupling constant $g_{a\gamma}$. Furthermore, the SNR scales approximately with the finesse and the cavity length. This dependence arises because a higher finesse increases the number of photons stored in the cavity, and a longer cavity extends the interaction length with the ALP field. Setting SNR=1 defines the sensitivity curve in the $g_{a\gamma}$ vs. $m_a$ parameter space.

\begin{figure}[H]
 \begin{tabular}{cc}
    (a) &  \\
     & \begin{minipage}[b]{0.9\linewidth}
         \centering
    \includegraphics[width=1.0\linewidth, bb = 0 0 360 224]{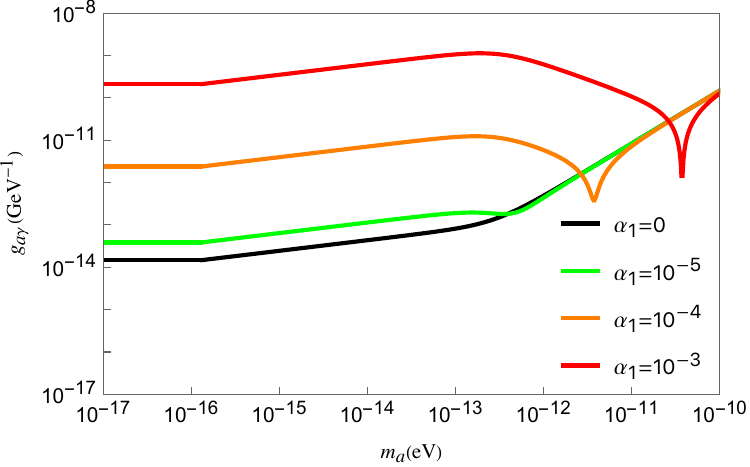}
     \end{minipage} \\
     (b) &  \\
      & \begin{minipage}[b]{0.9\linewidth}
          \centering
    \includegraphics[width=1.0\linewidth, bb = 0 0 360 224]{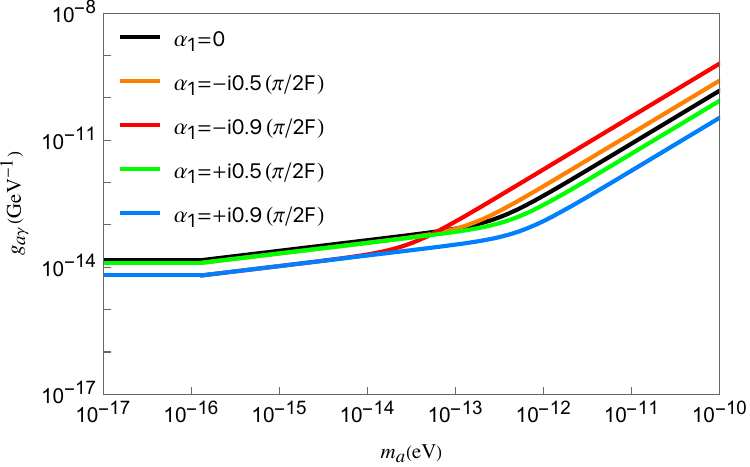}
      \end{minipage}
 \end{tabular}
    \caption{Sensitivity curves with (a) real and (b) imaginary retardation. The black lines are the cases without birefringence. We set $\theta_1=\theta_2=0$ so that the fast and slow axes of the waveplate coincide with polarizations of the carrier and signal light, respectively. The laser frequency is fixed at the resonance frequency of the carrier light, which means that the birefringence with real retardation effectively shifts the resonance curves for the sideband light. Degradations in (a) occur in the low-mass region because the resonance frequency is shifted from on-resonance to off-resonance by the birefringence. Dips in (a) are located at $m_aL=\mathrm{Re}[\alpha_1]$, where both carrier and signal light resonate in the cavity. In (b), the imaginary part of $\alpha_1$ is limited by (\ref{eq: Im alpha limitation approx}) because the actual reflectance $r_1e^{\pm\mathrm{Im}[\alpha_1]}$ is smaller than $1$. A nonzero $\mathrm{Im}[\alpha_1]$ splits reflectance, and, hence, the finesse is  also split; a positive (negative) $\mathrm{Im}[\alpha_1]$ enlarges the finesse of carrier (signal) light as in (\ref{eq: actual finesses}). Common parameters are $\mathcal{F}=10^5~(r_1=r_2)$, $L=10.64~\mathrm{m}$, $\lambda=1064~\mathrm{nm}$, $P=1~\mathrm{W}$, $T_\mathrm{obs}=1~\mathrm{yr}$, $\alpha_2=\alpha_1$, $\theta_1=\theta_2=0$, and $\epsilon=0.01~\mathrm{rad}$. The resonance condition in (\ref{eq: resonance condition theta_1,2=0}) holds.}
    \label{fig: sensitivity optimistic}
\end{figure}

Figure~\ref{fig: sensitivity optimistic} shows how sensitivity is affected by the birefringence of the mirrors. The parameters used for the calculation are the averaged finesse $\mathcal{F}=10^5$~(isotropic reflectance $r_1=r_2$), cavity length $L=10.64~\mathrm{m}$, laser wavelength $\lambda=1064~\mathrm{nm}$, laser power $P=1~\mathrm{W}$, observation time $T_\mathrm{obs}=1~\mathrm{yr}$, and postselection angle $\epsilon=0.01~\mathrm{rad}$. The results for $\alpha_1=0$ shown in Fig.~\ref{fig: sensitivity optimistic} are consistent with the calculation in \cite{Obata2018} aside from the parameter difference. The sensitivity decreases as the ALP mass increases in $m_a\gtrsim10^{-16}~\mathrm{eV}$ because of the time improvement factor in (\ref{eq: time improvement factor}), and the slope becomes steeper in $m_a\simeq10^{-13}\mathrm{eV}(\sim1/\mathcal{F}L)$ because the sideband becomes off-resonant.

For the parameters related to birefringence, we set $\alpha_1=\alpha_2$ and $\theta_1=\theta_2=0$ assuming that the mirrors are the same and that alignment is perfect. The laser frequency is fixed at the resonance frequency of the carrier light. Thus, the split of resonance curves by the birefringence with real retardation is translated as the shift of the resonance curves for the signal light. As shown in Fig.~\ref{fig: sensitivity optimistic}(a), the degradation begins when the retardation $\mathrm{Re}[\alpha]$ becomes greater than the inverse of the finesse $1/\mathcal{F}$, because the resonance frequency is shifted and the sideband light exits the resonance peak. However, the sensitivity becomes higher than that without birefringence when the sidebands become resonant because of the compensation for the frequency shift by the ALP mass; $2m_aL=\mathrm{Re}[\alpha]$. This simultaneous resonance is indicated by the dips shown in Fig.~\ref{fig: sensitivity optimistic}(a). When $\alpha_1=0$ and $\alpha_2\neq0(\in\mathbb{R})$, the degradation is halved because SNR depends mainly on $\alpha$, which is equal to $(\alpha_1+\alpha_2)$ when $\theta_1=\theta_2$. Fig.~\ref{fig: sensitivity optimistic}(b) shows the case of $\alpha_1\in i\mathbb{R}$, which causes a difference in the actual finesses $\mathcal{F}_{F/S}$ in (\ref{eq: actual finesses}). The effect of $\mathrm{Im}[\alpha_1]$ is not as large as that of $\mathrm{Re}[\alpha_1]$ because the values of $\mathrm{Im}[\alpha_{1,2}]$ are limited by (\ref{eq: Im alpha limitation}) or (\ref{eq: Im alpha limitation approx}). The sensitivity primarily follows the larger finesse, because the shot noise is governed by the total intracavity power, which is dominated by the higher-finesse mode. According to (\ref{eq: aprrox actual finesse eta_1}), this mode deviates more markedly from the non-birefringent limit than the lower-finesse mode. For $\mathrm{Im}[\alpha_1]>0$, the carrier light is enhanced, which results in a higher sensitivity in the entire mass region. For $\mathrm{Im}[\alpha_1]<0$, the signal light is enhanced, which causes higher sensitivity in the low-mass region where the sideband is on-resonant. However, because the sideband is completely off-resonant in the high-mass region, the effect of larger finesse does not appear and, instead, the effect of the smaller finesse of the carrier light appears.

\begin{figure}[H]
    \centering
    \includegraphics[width=1.0\linewidth, bb = 0 0 365 227]{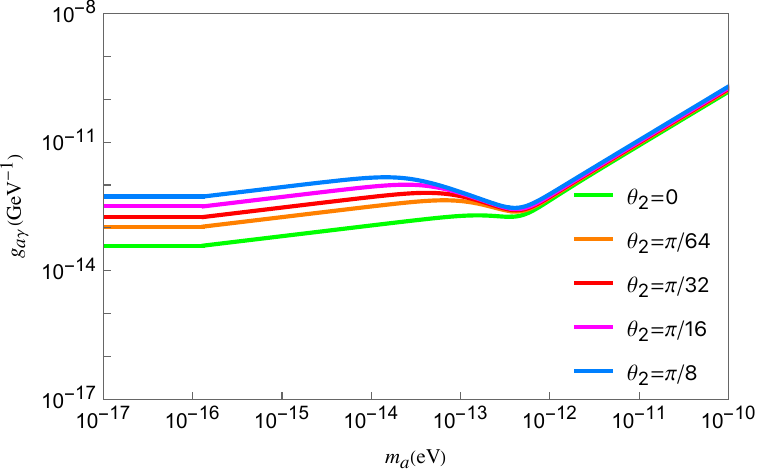}
    \caption{Sensitivity curves with finite values of $\theta_2$. The light green line is the same line as that shown in Fig. \ref{fig: sensitivity optimistic}(a) in which birefringence is relatively small and the sidebands are almost on resonance in the low-mass region. Degradation is enhanced when the ALP mass is lower than $\pi/2\mathcal{F}L$, where the resonance of signal light in the cavity is relevant. For finite $\theta_1$ and/or $\theta_2$, birefringence modifies linear polarizations to slightly elliptic ones. Roughly speaking, a finite $\theta_{1,2}$ makes the carrier and signal light flow to the detection and reference ports, respectively. Common parameters are $\mathcal{F}=10^5~(r_1=r_2)$, $L=10.64~\mathrm{m}$, $\lambda=1064~\mathrm{nm}$, $P=1~\mathrm{W}$, $T_\mathrm{obs}=1~\mathrm{yr}$, $\alpha_1=\alpha_2=10^{-5}~\mathrm{rad}$, and $\theta_1=0$. The resonance condition in (\ref{eq: resonance condition theta_1,2=0}) holds.}
    \label{fig: sensitivity theta}
\end{figure}

For fixed nonzero values of $\alpha_{1,2}$, the sensitivity also decreases as $\theta_{1,2}$ increases. Fig.~\ref{fig: sensitivity theta} shows the case in which $\theta_1=0$ and $\theta_2\neq0$. Degradation occurs in the low-mass region $m_a\ll \pi/2\mathcal{F}L$, whereas it does not occur in the high-mass region $m_a\gg \pi/2\mathcal{F}L$. In general, nonzero values of $\alpha_{1,2}$ and $\theta_{1,2}$ cause a flow of light to the other polarization. Fig.~\ref{fig: sensitivity theta} shows that only the flow of the signal light resonating in the cavity is relevant to sensitivity degradation because the signal light is close to the resonance in the low-mass region and far from the resonance in the high-mass region, because the flow of the carrier light to the detection port increases the shot noise in the same way; roughly speaking, the signal power is a multiplication of the amplitude of the signal light and that of the carrier light, whereas the square root of the shot noise power is proportional to the amplitude of carrier light. Thus, the flow of carrier light scarcely affects the sensitivity, whereas the flow of signal light to the reference port decreases the sensitivity. The degradation depends mainly on $\theta$, which is the ``average" of $\theta_{1,2}$, written as (\ref{eq: theta}).

\begin{figure}[H]
    \centering
    \includegraphics[width=1.0\linewidth, bb = 0 0 360 232]{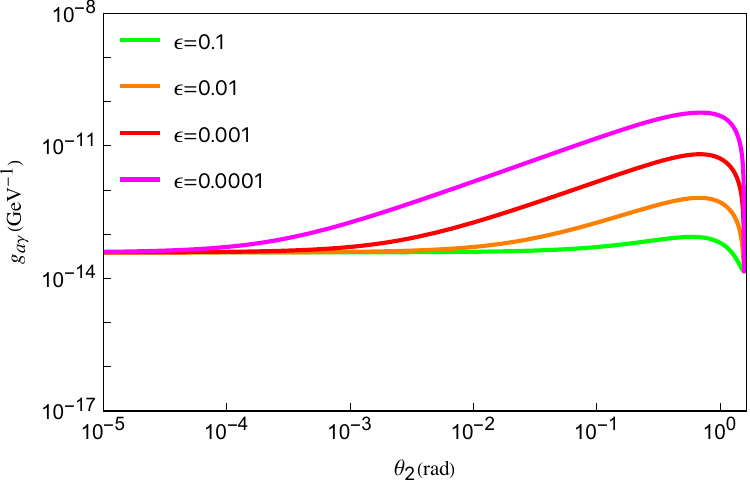}
    \caption{Dependence of sensitivity on the angle of birefringence axis $\theta_2$ and postselection angle $\epsilon$. Degradations occur when the angle $\theta_2$ is comparable to the postselection angle $\epsilon$. Conversely, degradation can be suppressed by taking a sufficiently large $\epsilon$. The sensitivity is symmetric with respect to $\theta_2=\pi/4$. Common parameters are $m_a=10^{-16}~\mathrm{eV}$, $\mathcal{F}=10^5~(r_1=r_2)$, $L=10.64~\mathrm{m}$, $\lambda=1064~\mathrm{nm}$, $P=1~\mathrm{W}$, $T_\mathrm{obs}=1~\mathrm{yr}$, $\alpha_1=\alpha_2=10^{-5}~\mathrm{rad}$, and $\theta_1=0$. The resonance condition in (\ref{eq: resonance condition theta_1,2=0}) holds.}
    \label{fig: sensitivity theta epsilon}
\end{figure}

Figure~\ref{fig: sensitivity theta epsilon} clearly shows that the sensitivity starts to degrade when $\theta_2$ is comparable to the postselection angle $\epsilon$. This result means that even if $\theta_{1,2}$ are nonzero, degradation can be suppressed by taking a sufficiently large $\epsilon$. The sensitivity worsens to $\theta_2=\pi/4$ and recovers to $\theta_2=\pi/2$, where the Jones matrix $R_2$ is equal to that with parameters $\alpha_2=-10^5$ and $\theta_2=0$.

\begin{figure}[h]
    \centering
    \includegraphics[width=\columnwidth]{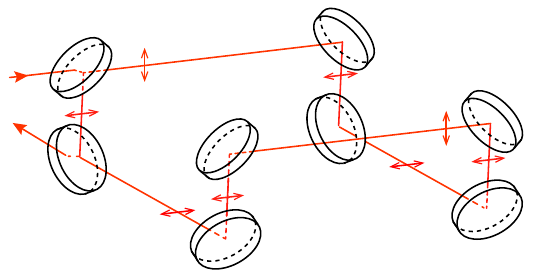}
    \caption{Conceptual design of the three-dimensional cavity. It comprises a ring cavity composed of eight mirrors, where the light path includes vertical segments as well as those in the horizontal plane. The light is input from the top-left mirror with vertical polarization in this figure. }
    \label{fig: 3dcavity}
\end{figure}

Although the eﬀects of birefringence can be suﬃciently mitigated by postselection, it is still worth considering a hardware-level birefringence suppression strategy. For example, dynamic variations in birefringence, especially when occurring at frequencies comparable to those of the dark-matter field, may still pose a challenge.
The use of an auxiliary cavity \cite{Martynov2020,Fujimoto2021}, zero-phase shift mirrors, and a wavelength tunable laser \cite{Takidera2025} has been proposed to solve this problem. Here, we propose a cavity design, as illustrated in Fig.~\ref{fig: 3dcavity}, which suppresses the intrinsic birefringence.
The s-polarized light incident on one mirror becomes p-polarized at the next mirror, and vice versa. This alternation cancels the relative phase shift after two reflections, effectively eliminating birefringence. A conceptual design and mode analysis are provided in Appendix F, where we show that this approach can achieve nearly circular cavity modes using only simple planar and spherical mirrors. Although a detailed hardware implementation is beyond the scope of this study, this concept offers a promising route toward birefringence-suppressed ALP search cavities.

\section{Conclusion\label{sec:conclusion}}
The birefringence of the optical components is a critical factor affecting the performance of optical cavities in precision experiments.
In this study, we developed an analytical model to describe the effects of birefringent mirrors on the polarization and resonance conditions in optical cavities.
We then applied this model to search for ALPs and quantitatively evaluated its impact on the measurement sensitivity.

In our model, a birefringent mirror is treated as a combination of an isotropic mirror and a waveplate characterized by complex retardation. This birefringence splits the resonance conditions of the two orthogonal linear-polarization modes. We first analyzed the mode mixing induced solely by birefringence without the ALP field.

In the context of ALP searches, the splitting of resonance peaks caused by birefringence degrades the sensitivity in the low-mass region owing to the shift of resonance frequency for signal light. However, in the high-mass region, where the ALP mass compensates for the resonance frequency shift caused by birefringence, the sidebands become resonant.
In such cases, the sensitivity is enhanced and can even surpass that of an ideal case without birefringence.
These findings are consistent with those of previous studies, and our more detailed treatment further confirms their validity.

We also showed that postselection plays a crucial role in mitigating the sensitivity degradation in the low-mass regions. By selecting an appropriate polarization at the detection port, the effect of birefringence-induced leakage can be suppressed. By contrast, the effect of postselection is negligible in the high-mass region, even near the enhanced sensitivity peaks resulting from simultaneous resonance.

ALP searches using high-finesse optical cavities remain a promising approach for probing ultralight pseudoscalar dark matter.
Our work provides a thorough analysis of the birefringence effect in such experiments, demonstrating that although birefringence can degrade sensitivity, this limitation can be overcome through careful cavity design and postselection techniques.

\begin{acknowledgements}
We thank Professor Noboru Sasao for stimulating this study.
We also thank Seinosuke Fujita for measuring the birefringence of mirrors as a starting point for this study. This work was supported by the Foundation of Kinoshita Memorial Enterprise and JST ASPIRE (JPMJAP2339). This work was also supported in part by a JSPS Kakenhi Grant No. JP19H05606.
\end{acknowledgements}

\appendix

\section{Interaction between ALP field and photons\label{sec:interaction}}

The interaction Lagrangian between ALPs and photons is expressed as
\begin{align}
    \mathcal{L}_{\mathrm{int}}=-\frac{1}{4}g_{a\gamma}aF_{\mu\nu}\tilde{F}^{\mu\nu},
\end{align}
where $g_{a\gamma}$ is the coupling constant between the ALP and photons, $F_{\mu\nu}=\partial_\mu A_\nu-\partial_\nu A_\mu$ is the field strength of the photon, and $\tilde{F}^{\mu\nu}=(1/2)\varepsilon^{\mu\nu\rho\sigma}F_{\rho\sigma}$ is its dual tensor. In the radiation gauge, the space component of the Euler--Lagrange equation of motion is
\begin{align}
    \left( \frac{\partial^2}{\partial t^2}-{\bf \nabla}^2 \right)\bm{A}=-g_{a\gamma}\dot{a}\left( \bm{\nabla} \times \bm{A} \right).
\end{align}
The ALP field is written as a classical field:
\begin{align}
    a(t)=a_0\cos{[m_at+\delta_\tau(t)]},
\end{align}
where $m_a$ is the ALP mass. The phase $\delta_\tau(t)$ can be regarded as a constant within the coherent time scale of ALP dark matter $\tau$, $\delta_\tau(t)=\delta_a$. The amplitude $a_0$ is related to the energy density of axion dark matter $\rho_{\mathrm{DM}}$ as $\rho_{\mathrm{DM}}=a_0^2m_a^2/2$.

We assume that the circularly polarized plane-wave solution along the $z$-axis has a time-dependent amplitude:
\begin{align}
    \bm{A}(z,t)=A(t)\frac{1}{\sqrt{2}}\begin{pmatrix} 1 \\ \pm i \end{pmatrix}e^{ikz},
\end{align}
where the basis of the two-vector representation is linear polarization; the upper and lower components correspond to horizontal and vertical linear polarizations, respectively. $A(t)$ follows
\begin{align}
 \label{eq:A(t)}
    \left(\frac{d^2}{dt^2}-k^2\right)A(t)=\mp 2gkm_a\sin{\left(m_at+\delta_a\right)}A(t),
\end{align}
where
\begin{align}
    \label{eq: perturbation parameter}
    g=g_{a\gamma}\sqrt{\rho_{\mathrm{DM}}/2}/m_a
\end{align}
is a dimensionless quantity with a typical value of
\begin{align}
 \begin{split}
    g\sim &1.1\times10^{-11}\left(\frac{g_{a\gamma}}{10^{-12}~\mathrm{GeV}^{-1}}\right) \\
    &\times\left(\frac{\rho_{\mathrm{DM}}}{0.3~\mathrm{GeV/cm^3}}\right)^{1/2}\left(\frac{10^{-13}~\mathrm{eV}}{m_a}\right).
 \end{split}
\end{align}
Because $g$ is significantly smaller than 1, (\ref{eq:A(t)}) can be perturbatively solved for arbitrary orders of $g$. Assuming that
\begin{align}
    A(t)&=A_0c_g(t)e^{-i\omega_gt}, \\
    c_g(t)&=1+gc_1(t)+g^2c_2(t)+\cdots, \\
    \omega_g&=k+g\omega_1+g^2\omega_2+\cdots,
\end{align}
with the initial conditions $c_g(0)=1$ and $k\gg m_a$, the solution up to $\mathcal{O}(g)$ is
\begin{align}
 \begin{split}
    c_g(t)&\simeq1\mp\frac{i}{2}g\bigl[\left(e^{-im_at}-1\right)e^{-i\delta_a} \\
    &\hspace{50pt}+\left(e^{im_at}-1\right)e^{i\delta_a} \bigr],
 \end{split} \\
    \omega_g&\simeq k,
\end{align}
where we assume that $k=2\pi/\lambda$ is the wave number of the laser with a wavelength of $\lambda\sim1000~\mathrm{nm}$, and we approximate
\begin{align}
    \frac{m_a}{k}\sim10^{-13}\left(\frac{m_a}{10^{-13}~\mathrm{eV}}\right)\left(\frac{\lambda}{1000~\mathrm{nm}}\right)
\end{align}
as zero.

The modification of the plane wave traveling along the length of $L$ can be interpreted as the time-evolution operator, which is given by 
\begin{align}
 \label{eq: ALP operator}
 \begin{split}
    U(t,t-L)&=1 \\
    &\hspace{10pt}-\frac{i}{2}g\bigl[\left(1-e^{im_aL}\right)e^{-i(m_at+\delta_a)} \\
    &\hspace{30pt}+\left(1-e^{-im_aL}\right)e^{i(m_at+\delta_a)} \bigr]\sigma_y,
 \end{split}
\end{align}
where $\sigma_y$ denotes the Pauli matrix.

\section{Details of state of light with ALP field}

The coefficient vectors in (\ref{eq: bmC^pm_f}) can be expressed as
\begin{align}
    \label{eq: u^pm_F}
    \bm{u}^{(\pm)}_F&:=\bm{b}_F+\frac{i\Xi_\pm-\Sigma_\pm}{1-e^{\pm2im_aL}}\bm{a}_F-\frac{\Sigma_\pm}{1-e^{\pm2im_aL}e^{-2i\alpha}}\bm{a}_S, \\
    \label{eq: u^pm_S}
    \bm{u}^{(\pm)}_S&:=\bm{b}_S+\frac{i\Xi_\pm+\Sigma_\pm}{1-e^{\pm2im_aL}}\bm{a}_S+\frac{\Sigma_\pm}{1-e^{\pm2im_aL}e^{2i\alpha}}\bm{a}_F, \\
    \label{eq: v^pm_F}
    \bm{v}^{(\pm)}_F&:=\frac{-1}{1-e^{\pm2im_aL}}\left[i\Xi_\pm+\frac{e^{\pm2im_aL}\left(e^{2i\alpha}-1\right)}{1-e^{\pm2im_aL}e^{2i\alpha}}\Sigma_\pm\right]\bm{a}_F, \\
    \label{eq: v^pm_S}
    \bm{v}^{(\pm)}_S&:=\frac{-1}{1-e^{\pm2im_aL}}\left[i\Xi_\pm-\frac{e^{\pm2im_aL}\left(e^{-2i\alpha}-1\right)}{1-e^{\pm2im_aL}e^{-2i\alpha}}\Sigma_\pm\right]\bm{a}_S,
\end{align}
where $\Xi_\pm$ is given in (\ref{eq: Xi_pm}), and
\begin{align}
    \Sigma_\pm&:=\frac{1}{2}\left[\left(\frac{\partial}{\partial\alpha}\Gamma\right)_{\alpha=0},\Xi_\pm\right], \\
    \bm{a}_{F/S}&:=e^{-i\theta\bm{n}\cdot\bm{\sigma}}\frac{1\pm\sigma_z}{2}e^{i\theta\bm{n}\cdot\bm{\sigma}}T_1\begin{pmatrix} 1 \\ 0 \end{pmatrix}, \\
    \bm{b}_{F/S}&:=e^{-i\theta\bm{n}\cdot\bm{\sigma}}\frac{1\pm\sigma_z}{2}e^{i\theta\bm{n}\cdot\bm{\sigma}}\left(-i\sigma_y\right)T_1\begin{pmatrix} 1 \\ 0 \end{pmatrix},
\end{align}
where $\sigma_y$ in $\Xi_\pm,\Sigma_\pm,\bm{b}_{F/S}$ originates from the time-evolution operator in (\ref{eq: ALP operator}). Therefore, all terms described by $\bm{u}_{F/S}^{(\pm)},\bm{v}_{F/S}^{(\pm)}$ are transited to the frequency $\omega_0\pm m_a$ by the ALP field.

\section{Simplest case: No polarization rotation at the mirrors}
\subsection{Without ALP field}

When $\theta_1=\theta_2=0$, the polarization does not rotate at the mirrors. This phenomenon implies no vertical polarization in the cavity when the effect of the ALP field is neglected. In this case, $\theta=0$, and $T_{1,2}$ is diagonal. Therefore,
\begin{align}
    a'_F&=t_1e^{\beta_1}t_2e^{\beta_2}\cos{\epsilon}, \\
    a'_S&=0.
\end{align}
The power in (\ref{eq: P'_f}) becomes
\begin{align}
    P'_f(t;\omega_0)\propto\left(t_1e^{\beta_1}t_2e^{\beta_2}\right)^2\cos^2{\epsilon}\left|\frac{1}{1-r_1r_2e^{2i\omega_0L}e^{-i\alpha}}\right|^2,
\end{align}
where $\alpha=\alpha_1+\alpha_2$. The first factor $\left(t_1e^{\beta_1}t_2e^{\beta_2}\right)^2$ is the transmittance of the mirrors for the horizontally polarized light. The second factor is the choice of the observed polarization. The last factor forms a Lorentz distribution around the resonance frequency, which satisfies the resonance condition:
\begin{align}
    \label{eq: resonance condition theta_1,2=0}
    2\omega_0L-\mathrm{Re}[\alpha]=2\pi l~~~(l\in\mathbb{Z}).
\end{align}
At approximately $\omega_0=\mathrm{Re}\left[\alpha\right]/2L~(l=0)$,
\begin{align}
    &\left|\frac{1}{1-r_1r_2e^{2i\omega_0L}e^{-i\alpha}}\right|^2 \\
    &\simeq \left|\frac{1}{1-r_1r_2e^{\mathrm{Im}\left[\alpha\right]}+ir_1r_2e^{\mathrm{Im}\left[\alpha\right]}\left(2\omega_0L-\mathrm{Re}\left[\alpha\right]\right)}\right|^2 \\
    &=\left(\frac{1}{1-r_1r_2e^{\mathrm{Im}\left[\alpha\right]}}\right)^2\frac{1}{1+\left[\frac{2}{\pi}\mathcal{F}_FL\left(\omega_0-\frac{\mathrm{Re}\left[\alpha\right]}{2L}\right)\right]^2},
\end{align}
where 
\begin{align}
    \mathcal{F}_F=\frac{\pi\sqrt{r_1e^{\mathrm{Im}\left[\alpha_1\right]}r_2e^{\mathrm{Im}\left[\alpha_2\right]}}}{1-r_1e^{\mathrm{Im}\left[\alpha_1\right]}r_2e^{\mathrm{Im}\left[\alpha_2\right]}}
\end{align}
is the finesse of the horizontally polarized light. This is a special case of (\ref{eq: actual finesses}).

\subsection{With ALP field}

Because the carrier light is purely horizontally polarized and the ALP field changes the polarization of the light, the signal light is purely vertically polarized. In other words, the carrier and signal light resonate only along the fast and slow axes of the waveplate, respectively. In this case,
\begin{align}
    \bm{u}_F^{(\pm)}&=\bm{0}, \\
    \bm{u}_S^{(\pm)}&=-t_1e^{\beta_1}\frac{e^{\pm im_aL}\left(e^{2i\alpha_1}+e^{\pm im_aL}e^{2i\alpha}\right)}{1-e^{\pm2im_aL}e^{2i\alpha}}\begin{pmatrix} 0 \\ 1 \end{pmatrix}, \\
    \bm{v}_F^{(\pm)}&=t_1e^{\beta_1}\frac{1+e^{\pm m_aL}e^{2i\alpha_1}}{1-e^{\pm2im_aL}e^{-2i\alpha}}\begin{pmatrix} 0 \\ 1 \end{pmatrix}, \\
    \bm{v}_S^{(\pm)}&=\bm{0},
\end{align}
and
\begin{align}
 \label{eq: C_f^pm theta_1,2=0}
 \begin{split}
    \bm{C}_f^{(\pm)}&=\frac{t_1e^{\beta_1}}{1-e^{\pm2im_aL}e^{2i\alpha}}\Biggl[\frac{1+e^{\pm m_aL}e^{2i\alpha_1}}{1-r_1r_2e^{2i\omega_0L}e^{-i\alpha}} \\
    &~~~~~~-\frac{e^{\pm im_aL}\left(e^{2i\alpha_1}+e^{\pm im_aL}e^{2i\alpha}\right)}{1-r_1r_2e^{2i(\omega_0\pm m_a)L}e^{i\alpha}}\Biggr]\begin{pmatrix} 0 \\ 1 \end{pmatrix}.
 \end{split}
\end{align}
In this equation, two of the four factors vanish and the $\bm{C}_f^{(\pm)}$ is vertically polarized.

\section{Approximate resonance condition}

When $\theta_{1,2}=0$, the resonance condition is (\ref{eq: resonance condition theta_1,2=0}). In the general case of $\theta_{1,2}\neq0$, the distribution becomes a ``superposition" of two Lorentzians with the two different resonance conditions in (\ref{eq: resonance condition}). Therefore, the resonance condition for the overall distribution $P'_f$ may not be accurately determined, and the distribution can have two peaks. This ambiguity arises when the shift in the peaks becomes larger than their widths $\mathrm{Re}[\alpha]\gtrsim\pi/\mathcal{F}_{F/S}$. Even when $\mathrm{Re}[\alpha]\ll\pi/\mathcal{F}_{F/S}$, the resonance condition is difficult to derive because the distribution is no longer Lorentzian. To obtain the approximate resonance condition, we assume that the imaginary part of the combined retardation $\alpha$ is small. More precisely, we impose the following condition:
 \begin{align}
  \label{eq: small alpha}
     |\alpha|\ll\frac{\pi}{\mathcal{F}}.
 \end{align}
This condition allows us to approximate the overall distribution as a single Lorentzian. Under this condition, the resonance factor can be approximated as
\begin{align}
    \frac{1}{1-r_1r_2e^{2i\omega_0L}e^{\mp i\alpha}}&\simeq\frac{1}{1-r_1r_2e^{2i\omega_0L}\pm i\alpha r_1r_2e^{2i\omega_0L}} \\
    &=\frac{1}{1-r_1r_2e^{2i\omega_0L}}\frac{1}{1\pm i\alpha \frac{r_1r_2e^{2i\omega_0L}}{1-r_1r_2e^{2i\omega_0L}}} \\
    \begin{split}
     &\simeq\frac{1}{1-r_1r_2e^{2i\omega_0L}} \\
     &~~~~~~\times\left(1\mp i\alpha \frac{r_1r_2e^{2i\omega_0L}}{1-r_1r_2e^{2i\omega_0L}}\right),
    \end{split}
\end{align}
and the power is
\begin{align}
 \begin{split}
    P'_f(t;\omega_0)&\propto\Biggl|\frac{1}{1-r_1r_2e^{2i\omega_0L}}\Biggl[\left(a'_F+a'_S\right) \\
    &~~~~~~~-i\alpha\left(a'_F-a'_S\right)\frac{r_1r_2e^{2i\omega_0L}}{1-r_1r_2e^{2i\omega_0L}}\Biggr]\Biggr|^2
    \end{split} \\
    &\simeq\left|\frac{a'_F+a'_S}{1-r_1r_2e^{2i\omega_0L}e^{-i\alpha\frac{a'_F-a'_S}{a'_F+a'_S}}} \right|^2
\end{align}
In the final approximation, we assume that $|(a'_F-a'_S)/(a'_F+a'_S)|\sim1$. This is a single Lorentzian with the following resonance condition:
\begin{align}
    2\omega_0L-\mathrm{Re}\left[\alpha\frac{a'_F-a'_S}{a'_F+a'_S}\right]=2\pi l~~~(l\in\mathbb{Z}).
\end{align}
To determine $(a'_F-a'_S)/(a'_F+a'_S)$, we restrict the original parameters $\alpha_{1,2},\theta_{1,2}$; $\alpha_1=\alpha_2$ and $\theta_1=\theta_2$ for simplicity. The combined parameters are
\begin{align}
    \alpha&=\alpha_1+\alpha_2=2\alpha_1 \\
    \theta&=\theta_1=\theta_2 \\
    \phi&=0
\end{align}
and the condition in (\ref{eq: small alpha}) becomes $|\alpha_{1,2}|\ll\pi/2\mathcal{F}$. These calculations are performed using linear approximation of $\alpha$ and $\alpha(a'_F-a'_S)/(a'_F+a'_S)$. We consider $(a'_F-a'_S)/(a'_F+a'_S)$ with $\alpha=0$ or $\alpha_1=\alpha_2=0$. The denominator is
\begin{align}
    \left(a'_F+a'_S\right)_{\alpha=0}&=\bm{n}'_f\cdot t_2e^{i\theta\bm{n}\cdot\bm{\sigma}}\cdot1\cdot e^{-i\theta\bm{n}\cdot\bm{\sigma}}t_1\begin{pmatrix} 1 \\ 0 \end{pmatrix} \\
    &=t_1t_2\cos{\epsilon},
\end{align}
and the numerator is
\begin{align}
    \left(a'_F+a'_S\right)_{\alpha=0}&=\bm{n}'_f\cdot t_2e^{i\theta\bm{n}\cdot\bm{\sigma}}\cdot\sigma_z\cdot e^{-i\theta\bm{n}\cdot\bm{\sigma}}t_1\begin{pmatrix} 1 \\ 0 \end{pmatrix} \\
    &=t_1t_2\bm{n}'_f\cdot\begin{pmatrix} \cos{(2\theta)} & \sin{(2\theta)} \\ \sin{(2\theta)} & \cos{(2\theta)} \end{pmatrix}\begin{pmatrix} 1 \\ 0 \end{pmatrix} \\
    &=t_1t_2\left(\cos{\epsilon}\cos{(2\theta)}-\sin{\epsilon}\sin{(2\theta)}\right) \\
    &=t_1t_2\cos{(2\theta+\epsilon)}.
\end{align}
Thus,
\begin{align}
    \frac{a'_F-a'_S}{a'_F+a'_S}=\frac{\cos{(2\theta+\epsilon)}}{\cos{\epsilon}}
\end{align}
is at least an order-one quantity when $\alpha_1=\alpha_2$ and $\theta_1=\theta_2$. The resonance condition is
\begin{align}
    \label{eq: resonance condition approx 1=2}
    2\omega_0L-\mathrm{Re}\left[\alpha\right]\frac{\cos{(2\theta+\epsilon)}}{\cos{\epsilon}}=2\pi l~~~(l\in\mathbb{Z}).
\end{align}

\section{Translation of measurable quantities into parameters of mirrors}

In a simple case, in which the two mirrors are the same and the alignment is perfect, the values of the birefringence parameters can be determined from the measurable quantities. In this case, we can reduce the number of parameters by
\begin{align}
    \alpha_1&=\alpha_2=2\alpha, \\
    \theta_1&=\theta_2=0,
\end{align}
and
\begin{align}
    r_1=r_2=:r~\left(\mathcal{F}=\frac{\pi\sqrt{r_1r_2}}{1-r_1r_2}=\frac{\pi r}{1-r^2}\right).
\end{align}
Here, the number of unknown parameters is three: $\mathrm{Re}[\alpha_1]$, $\mathrm{Im}[\alpha_1]$, and $r$ (or $\mathcal{F}$).The resonance curves of the signal power can be measured by inserting lights with two polarizations. From the results, the values of detuning of the resonance frequencies between polarizations $\Delta f_\mathrm{res}$ and finesses along the two polarizations $\mathcal{F}_{F/S}$ can be defined.

The real part of the retardation $\mathrm{Re}[\alpha_1]$ can be determined using $\Delta f_\mathrm{res}$. From (\ref{eq: resonance condition}),
\begin{align}
    2\left(2\pi f_\mathrm{res}^+\right)L-2\mathrm{Re}[\alpha_1]=2\pi l
\end{align}
and
\begin{align}
    2\left(2\pi f_\mathrm{res}^-\right)L+2\mathrm{Re}[\alpha_1]=2\pi l,
\end{align}
where $\Delta f_\mathrm{res}=f_\mathrm{res}^+-f_\mathrm{res}^-$, and
\begin{align}
    \mathrm{Re}[\alpha_1]=\pi\Delta f_\mathrm{res}L.
\end{align}

The imaginary part of the retardation $\mathrm{Im}[\alpha_1]$ and average finesse can be determined using $\mathcal{F}_{F/S}$. From (\ref{eq: actual finesses}),
\begin{align}
    r^2\mathcal{F}_{F/S}e^{\pm 2\mathrm{Im}[\alpha_1]}+\pi re^{\pm \mathrm{Im}[\alpha_1]}-\mathcal{F}_{F/S}=0,
\end{align}
which are quadratic equations with respect to $e^{\pm \mathrm{Im}[\alpha_1]}$. The solution is
\begin{align}
    e^{\pm \mathrm{Im}[\alpha_1]}&=\frac{-\pi r+\sqrt{\pi^2r^2+4r^2\mathcal{F}_{F/S}^2}}{2r^2\mathcal{F}_{F/S}} \\
    &\simeq \frac{1}{r}\left(1-\frac{\pi}{2\mathcal{F}_{F/S}}\right),
\end{align}
where we assume $\mathcal{F}_{F/S}\gg 1$. Subsequently, the averaged reflectance is
\begin{align}
    r^2&=\left(re^{\mathrm{Im}[\alpha_1]}\right)\times\left(re^{-\mathrm{Im}[\alpha_1]}\right) \\
    &=\left(1-\frac{\pi}{2\mathcal{F}_F}\right)\left(1-\frac{\pi}{2\mathcal{F}_S}\right) \\
    &\simeq 1-\frac{\pi}{2}\frac{\mathcal{F}_F+\mathcal{F}_S}{\mathcal{F}_F\mathcal{F}_S},
\end{align}
or the average finesse is
\begin{align}
    \mathcal{F}&=\frac{\pi r}{1-r^2}=\frac{\pi\left(1-\frac{\pi}{4}\frac{\mathcal{F}_F+\mathcal{F}_S}{\mathcal{F}_F\mathcal{F}_S}\right)}{\frac{\pi}{2}\frac{\mathcal{F}_F+\mathcal{F}_S}{\mathcal{F}_F\mathcal{F}_S}} \\
    &\simeq \frac{2\mathcal{F}_F\mathcal{F}_S}{\mathcal{F}_F+\mathcal{F}_S}.
\end{align}
The imaginary part of the retardation is
\begin{align}
    e^{\mathrm{Im}[\alpha_1]}&=\frac{1}{r}\left(1-\frac{\pi}{2\mathcal{F}_F}\right) \\
    &\simeq \left(1+\frac{\pi}{4}\frac{\mathcal{F}_F+\mathcal{F}_S}{\mathcal{F}_F\mathcal{F}_S}\right)\left(1-\frac{\pi}{2\mathcal{F}_F}\right) \\
    &\simeq 1+\frac{\pi}{4}\frac{\mathcal{F}_F-\mathcal{F}_S}{\mathcal{F}_F\mathcal{F}_S}.
\end{align}
Hence, $\mathrm{Im}[\alpha_1]$ is significantly small, and
\begin{align}
    \mathrm{Im}[\alpha_1]\simeq\frac{\pi}{4}\frac{\mathcal{F}_F-\mathcal{F}_S}{\mathcal{F}_F\mathcal{F}_S}.
\end{align}

The value of $\mathrm{Im}[\alpha]$ is extremely small for a high finesse because the values of $\mathrm{Im}[\alpha_{1,2}]$ are limited by (\ref{eq: Im alpha limitation approx}). We can then approximate $\mathcal{F}_{F/S}$ with respect to $\mathrm{Im}[\alpha]$. We define $\eta$ as
\begin{align}
    \mathrm{Im}[\alpha]=\eta\frac{\pi}{2\mathcal{F}}.
\end{align}
The value of $\eta$ is limited by (\ref{eq: Im alpha limitation approx}). From (\ref{eq: actual finesses}), the approximations of $\mathcal{F}_{F/S}$ are
\begin{align}
    \label{eq: aprrox actual finesse eta}
    \mathcal{F}_{F/S}\simeq\frac{\pi r(1\pm\pi\eta/2\mathcal{F})}{1-r^2(1\pm\pi\eta/\mathcal{F})}\simeq\frac{\mathcal{F}}{1\mp\eta/2}.
\end{align}
By defining $\eta_{1,2}$ such that
\begin{align}
    \label{eq: def of eta_1,2}
    \mathrm{Im}[\alpha_{1,2}]=\eta_{1,2}\frac{\pi}{2\mathcal{F}},~~~~~~~(|\eta_{1,2}|<1),
\end{align}
the relation between $\eta$ and $\eta_{1,2}$ can be derived using (\ref{eq: alpha}) and is generally highly complicated. However, when $\theta_1=\theta_2$ and $\alpha_1=\alpha_2$, the relation becomes simple: $\eta=\eta_1+\eta_2=2\eta_1$ because $\alpha=\alpha_1+\alpha_2=2\alpha_1$ in this case. Then, we have
\begin{align}
    \label{eq: aprrox actual finesse eta_1}
    \mathcal{F}_{F/S}\simeq\frac{\mathcal{F}}{1\mp\eta_1},~~~~(|\eta_1|<1).
\end{align}
The second approximation in (\ref{eq: aprrox actual finesse eta}) is valid when $(1-|\eta|/2)$ is much larger than $\pi/\mathcal{F}$. Hence, (\ref{eq: aprrox actual finesse eta_1}) is valid when $(1-|\eta_1|)$ is significantly greater than $\pi/\mathcal{F}$.

\section{Conceptual design for birefringence suppression}

The design shown in Fig.~\ref{fig: 3dcavity} comprises a ring cavity composed of eight mirrors arranged in a three-dimensional structure, where the light path includes vertical segments as well as those in the horizontal plane. In conventional ring cavities constructed on a single plane, the s- and p-polarized components remain unchanged throughout a round trip; however, the polarization is altered at each mirror in the proposed design. This alternation cancels the relative phase shift after two reflections and eliminates the birefringence effectively. Moreover, because all mirrors can be coated identically, the temperature dependence of birefringence is expected to be uniform, allowing for the suppression of both static and dynamic birefringence effects.
As shown in the figure, the vertical light paths are short, whereas the horizontal paths are relatively long. Because the optical paths alternate between short and long segments, the rotation of polarization induced by ALPs, canceled under the assumption of a planar cavity as discussed in the main text, can remain unbalanced, thereby preserving sensitivity.

\begin{figure}[H]
    \centering
    \includegraphics[width=1.0\linewidth]{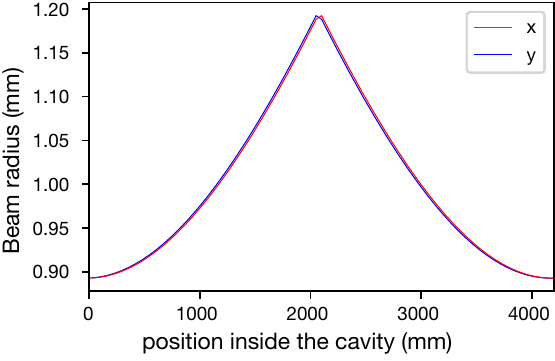}
    \caption{Round-trip eigenmode inside the cavity. The red line represents the beam-radius trace of a linear-polarization component shown in Fig.~\ref{fig: 3dcavity}, and the blue line represents the orthogonal polarization component.}
    \label{fig: 3dcavityb}
\end{figure}

A simple implementation uses parabolic mirrors; however, good performance can be achieved by placing two spherical mirrors with large radii of curvature in one vertical pair and using flat mirrors for the remaining segments. This configuration enables a nearly circular cavity mode with minimal aberration. For example, for a vertical path length of 5\,cm, horizontal side length of 1\,m, and radius of curvature of the spherical mirrors of 10\,m, the resulting cavity mode is shown in Fig.~\ref{fig: 3dcavityb}. The beam radius remains within 0.5\% in both transverse directions throughout the cavity, indicating an almost perfectly circular mode. By using only simple planar and spherical mirrors, imperfections in the mirror coatings can be minimized, thereby enabling a cavity system that approaches ideal performance. Although increasing the number of mirrors requires a more rigid and stable mechanical structure, this configuration presents an intriguing approach for birefringence-suppressed ALP dark-matter searches.

\bibliography{axion}
\end{document}